\documentstyle[12pt]{article}
\begin{document}
\title{Remarks on the spectral properties \\of tight binding
and Kronig-Penney models\\ with substitution sequences
\thanks{Work partially supported by the Commission of the European
Communities under contract No. SC1-CT91-0695}
}
\author{Anton Bovier \thanks{e-mail: bovier@iaas-berlin.d400.de}\\
Institut f\"ur Angewandte Analysis und Stochastik\\
Mohrenstrasse 39, D-10117 Berlin, Germany\\
and \\
Jean-Michel Ghez \thanks{e-mail: ghez@cpt.univ-mrs.fr}\\
Centre de Physique Th\'eorique - CNRS 	Luminy, Case 907
\\
	F-13288 Marseille Cedex 9, France \\
and	PHYMAT, D\'epartement de Math\'ematiques \\
	Universit\'e de Toulon et du Var, B.P. 132 \\
	F-83957 La Garde Cedex, France}
\date{18 January 1994}
\maketitle
\begin{abstract}
We comment on some  recent investigations on the electronic properties of
models associated to the Thue-Morse chain and point out that their conclusions
are in contradiction with rigorously proven theorems and indicate some
of the sources of these misinterpretations. We briefly review and explain the
current status of  mathematical results in this field and discuss some
conjectures and open problems.
\end{abstract}

\section{Introduction}

Following the discovery of quasicrystals by Shechtman et al. \cite{[1],[2]}
there has been a continuing interest of both physicists and mathematicians in
structures that exhibit what has been termed {\it deterministic disorder}. A
class of models that has attracted particular attention in this context are
one dimensional Schr\"odinger operators with potentials obtained from
so-called {\it substitution sequences} \cite{[3],[4]} and a number of
analytical and numerical tools have been developed for their investigation. On
the other hand, substitution sequences provide examples of various types of
aperiodic structures, e.g. quasiperiodic or not, that can be characterized by
the nature of their Fourier spectrum  which may be dense pure point (Fibonacci
sequence), singular continuous (Thue-Morse sequence), or even absolutely
continuous (Rudin-Shapiro sequence). It is naturally of great interest to
investigate the spectral and transport properties of systems in dependence of
such properties. Let us mention that models based on substitution sequences
have since enjoyed an increasing popularity also in different contexts such as
one-dimensional quantum Ising chains \cite{[5],[6]}, aperiodically kicked
quantum systems  \cite{[7]}, etc.

About a decade has passed since the pioneering papers by Kohmoto et al.
\cite{[8]} and Ostlund et al. \cite{[9]} appeared and a vast amount of
knowledge has since been accumulated through the work of both physicists and
mathematicians and through methods ranging from numerical simulations,
judicious guessing to heuristic and rigorous mathematics
\cite{[10],[11],[12],[13],[14],[15],[16],[17],[18],[19],[20],[21],[22]}.
In spite of these efforts, we are today still far from a complete and coherent
understanding of the properties of these systems which are, by and large, both
subtle and unusual. This situation, together with the prospective
technological applications (e.g. superlattices corresponding to substitution
sequences can today be manufactured \cite{[23],[24]}), invites a continued and
coordinated effort to further research in this area.

A regrettable feature of the current situation is that in spite of a
flourishing litterature and the existence of a number of review papers
\cite{[2],[25],[26]}, the communication between different schools has been
less than perfect and in particular it appears that what has been obtained in
terms of mathematically rigorous results has not generally been recognized. As
a result, there have appeared still very recently a number of papers in
reknowned journals (see in particular \cite{[27],[28],[29],[30]}) in which, on
the basis of numerical and heuristic methods, results are claimed that are in
striking contradiction to rigorously proven facts.

The present article is an attempt to improve this situation by explaining some
of the mathematical results concerning the spectral theory of Schr\"odinger
operators with substitution potentials and to rectify the errors in several
recent papers that came to our attention. More importantly, we will try to
explain the sources of the misinterpretations in these papers. Also, we would
like to indicate, in an informal way that should be accessible to
non-mathematicians, what mechanisms are important in leading to the
mathematical results, what further results may be expected and what
information would be needed to obtain these. We will also try to point to the
serious open problems in the field. To a lesser extend, this article may also
serve as a survey for the less involved reader.

Let us  briefly review the types of models we consider. First we recall the
definition of a {\it substitution sequence}. Take a finite set $\cal A$,
called an {\it alphabet} and denote by $\cal A^*$ the set of all
finitely long words that can be written in this alphabet. Moreover, we
write $\cal A^{\bf N}$ and ${\cal A}^{\bf Z}$ for the sets of all
semi-infinite and infinite sequences of letters from $\cal A$,
respectively. Now let $\xi$ be a map from $\cal A$ to $\cal A^*$, i.e. a rule
that associates to any letter in $\cal A$ a finite word. We call $\xi$
a {\it substitution rule} and extend it to a map from $\cal A^*$ to
$\cal A^*$ by specifying that $\xi$ acts on a word by substituting each letter
$\alpha_i$ of this word by its corresponding image $\xi(\alpha_i)$. By
the same rule the action of $\xi$ is extended to $\cal A^{\bf N}$ and
$\cal A^{\bf Z}$. A sequence $u\in \cal A^{\bf N}$ is then called
a {\it substitution sequence}, if it is a fixpoint of $\xi$, i.e. if it
remains invariant if each letter in the sequence is replaced by its image
under $\xi$. Some simple conditions on $\xi$ assure the existence of such
fixpoints. Moreover, except for trivial examples of substitution rules, this
fixpoints are aperiodic sequences. Examples of substitution sequences that
have attracted most attention in physics are

\begin{enumerate}
\item The Fibonacci sequence. Here ${\cal A}=\{a,b\}$ and the substitution
 rule is simply
\begin{equation}
a\rightarrow \xi(a)=ab,\quad b\rightarrow \xi(b)=a
\end{equation}
\item The Thue-Morse sequence. Again ${\cal A}=\{a,b\}$, but the rule is
this time
\begin{equation}
a\rightarrow \xi(a)=ab,\quad b\rightarrow \xi(b)=ba
\end{equation}
\item The period-doubling sequence. Again ${\cal A}=\{a,b\}$, and the rule is
\begin{equation}
a\rightarrow \xi(a)=ab,\quad b\rightarrow \xi(b)=aa
\end{equation}
\item The Rudin-Shapiro sequence. Here ${\cal A}=\{a,b,c,d\}$,
 and the rule is
\begin{equation}
a\rightarrow \xi(a)=ac,\quad b\rightarrow \xi(b)=dc,\quad c\rightarrow
\xi(c)=ab,\quad d\rightarrow \xi(d)=db
\end{equation}
\end{enumerate}

All of these examples, and in fact all substitutions we will consider  have
the property of being {\it `primitive'} \cite{[4]} which means, that there
exists an  integer $k$ such that for all pairs of letters
$\alpha,\beta$ in $\cal A$, the word $\xi^k(\alpha)$ contains the
letter $\beta$. Here $\xi^k$ means the $k$-fold application of $\xi$.
In the sequel the term substitution will always be understood to mean
`primitive substitution'.

Given a substitution sequence, one may consider various associated
Schr\"odinger operators. The most studied one is doubtlessly the tight binding
model, where the Hamiltonian act on wave functions $\psi$ in the Hilbert space
$\ell^2(\bf Z)$ of square integrable sequences as
\begin{equation}
H_{tb}\psi(n)\equiv \psi(n+1)+\psi(n-1)+ v_n\psi(n)
\end{equation}
Here the potential $v_n$ is obtained by assigning real value
$v(\alpha)$ to each letter in $\cal A$ and setting $v_n=v(\alpha)$, if
the $n$-th letter in the substitution sequence $u$ is $\alpha$.
Moreover, for negative values of $n$ one sets $v_n=v_{-n-1}$.

Another example are models of the Kronig-Penney type which have for instance
been proposed to describe transport in structured superlattices
\cite{[30],[31],[32],[33],[34],[35],[36],[37],[38],[39],[40]}. Here the
operator is defined on continuum wave functions in $L^2(\bf R)$ and is given
by
\begin{equation}
H_{KP}=-\frac {d^2}{dx^2} + V(x)
\end{equation}
where $V(x)$ is a step-function describing a sequence of potential barriers of
$|\cal A|$ types that are again arranged consecutively according to the chosen
underlying substitution sequence.

As always in the theory of differential operators one is interested in the
{\it spectrum} of the operators $H$. Let us recall that the spectrum,
$\sigma(H)$, of a self-adjoint operator $H$ is defined as the
complement of the set of values $E$ for which the {\it resolvent},
$(H-E)^ {-1}$ is a bounded operator. An important criteria that
characterizes the spectrum in the case of Schr\"odinger operators is
that $\sigma(H)$ coincides with the closure of the set of values $E$
for which the time independent Schr\"odinger equation
\begin{equation}
H\psi=E\psi
\label{eq1}
\end{equation}
possesses a solution that is polynomially bounded, i.e. for which there exist
constants $a$ and $b$ such that $|\psi_E(n)|\leq b |n|^a$, for all
$n\in \bf Z$. The values for which $\psi_E(n)$ is a square integrable
function are called the eigenvalues of $H$, and the closure of the set
of eigenvalues is called the point spectrum; the remaining spectrum is
the continuous spectrum which can be further decomposed into the
absolutely continuous and singular continuous parts, in accordance
with the Lebesgue decomposition of the spectral measure. Roughly
speaking, the absolutely continuous spectrum is a closed set with
non-empty interior,  while the singular continuous part (`what
remains' of the spectrum after the point and the absolutely continuous parts
have been removed) is  a Cantor set. The spectral type has important
consequences for the transport properties  of the models; if the Fermi-energy
falls in the absolutely continuous  part of the spectrum, we expect a
conductor, while if it falls into the point spectrum (or in the complement of
the spectrum) one expects to have an insulator. The singular continuous
spectrum has generally been considered to be somewhat exotic and is much less
understood; however, for the very models we are dealing with here it is quite
common and is expected to give rise to interesting new transport phenomena.

Basically since the early papers of Kohmoto et al. it had been conjectured
that, at least in the case of the Fibonacci sequence, the operator $H_{tb}$
should have singular continuous spectrum. This was later proven in two
remarkable papers by S\"ut\H o \cite{[12]} and (for more general Fibonacci
sequences) Bellissard et al. \cite{[13]}. It is a natural question to ask
whether this property depends on particular features of the Fibonacci
sequence, e.g. on the fact that it is quasi-periodic, and one may pose the
question whether and how the spectral type of the Hamiltonian reflects certain
features of the sequence, and in particular the nature of the Fourier spectrum
of the sequence, or if it could be an indicator for the degree of `randomness'
of the substitution sequence. The most prominent example of a substitution
sequence that is not quasi-periodic is certainly the Thue-Morse sequence and
this has led to a rather extensive investigation of this example. The question
of the spectral type in this case has been discussed in a number of papers
\cite{[14],[15],[16],[17],[18],[27],[28],[29],[30],[39]} both for the
tight-binding model and the Kronig-Penney model. In their most recent papers,
Ryu et al. \cite{[30]} arrive at the conclusion that in both cases the
spectrum contains an absolutely continuous component. On this basis they argue
that the Thue-Morse sequence should be regarded as more `periodic' than the
Fibonacci sequence. Unfortunately, their claim is false. In fact, in
\cite{[18]} (see also \cite{[17]} for a slightly incomplete argument) it has
been proven rigorously that the spectrum of the Thue-Morse model is purely
singular continuous (this paper deals mainly with another sequence, the period
doubling sequence, but the result applies also to the Thue-Morse case (see the
remark following Theorem 3 in \cite{[18]}). In fact, a much more general
result on the absence of absolutely continuous spectrum has been obtained in
\cite{[22]} which suggests that the spectral type is quite independent on the
particular properties of the substitution sequence. In the next sections we
explain those results in more detail and comment in more detail on how the
erroneous claims in \cite{[30]} have been obtained.

\vskip1cm
\section{Sequences and spectra}

It is a natural question to ask what properties of a given sequence determine
the spectral type of the corresponding Schr\"odinger operator. Classical
candidates might appear to be
\begin{enumerate}
\item the {\it entropy} of a sequence, defined as (see e.g. \cite{[4]})
$$
I=\lim_{k\uparrow\infty}\frac 1k\ln\#\left\{\hbox{different words of length
$k$ occurring in the sequence}\right\}
$$
\item the Fourier spectrum of the sequence.
\end{enumerate}

However, the entropy of a sequence is clearly too crude a measure; all
substitution sequences have zero entropy and are thus not distinguishable from
periodic ones. Nonetheless, their spectra are quite different. It is an
interesting, and as yet not much investigated question, whether more refined
quantities related to the number of different words in a sequence are related
to spectral properties. However, again, within the class of substitution
sequences this quantity always behaves in the same way and no distinction can
be seen.

The Fourier spectrum (see e.g. \cite{[41],[42]}), on the other hand, varies
widely in nature between substitution sequences, ranging all the way from pure
point (Fibonacci) over singular continuous (Thue-Morse) to absolutely
continuous (Rudin-Shapiro). One might naturally think that this should be
reflected in the spectra of the corresponding operators: pure point Fourier
should give rise to singular continuous spectrum of the Schr\"odinger operator
and absolutely continuous Fourier aught to correspond to point spectrum, with
the intermediate singular continuous Fourier somewhere in between.  However,
there is little evidence to support such conjectures. An attempt to provide a
theoretical basis to relate properties of the Fourier spectrum of a sequence
to spectral properties of the Schr\"odinger operators was made by Luck
\cite{[19]} on the basis of perturbation theory. Although his arguments are
mathematically heuristic, he succeeded in predicting the scaling of the
spectral gaps, at least in  cases where the Fourier spectrum is pure point. In
other cases, such as Thue-Morse and Rudin-Shapiro, there were no conclusive
predictions. This method does demonstrate, however, that a relation between
Fourier spectrum and the spectrum of the operator is basically limited to
first order perturbation theory, and higher order effects will become relevant
if the singularities in the Fourier spectrum are not strong enough.

A more pragmatic point of view would favour the idea that the crucial features
of a substitution sequence should reside in its self-similarity which is
encoded in its substitution rule. From that point of view, it would seem more
natural to conjecture that the spectral type for all substitution sequences
(excepting maybe some pathological cases) should be the same, namely singular
continuous spectrum. There are a number of sound mathematical results
indicating in this direction, with a number of open problems left to close the
argument and the remainder of this section is devoted to their explanation.

In all the one dimensional models of the type we consider here, i.e. the tight
binding or the Kronig-Penney models, the basic tool of spectral analysis is
the transfer matrix formalism. Without entering into the presentation of
details that have been presented many times elsewhere (for the Kronig-Penney
model see e.g. \cite{[30],[31],[32],[33],[34],[35],[36],[37],[38]}) we just
note that in all these cases this leads to the investigation of a product of
two-by two matrices of the form
\begin{equation}
P_n(E)=\prod_{k=1}^{n} T_E(u_{n-k})
\end{equation}
where $u_k$ is the $k$-th letter in the substitution sequence $u$ and
$T_E:{\cal A}\rightarrow SL(2,\bf C)$ is a map that assigns, for fixed
energy $E$, to each letter in the alphabet a unimodular two-by-two
matrix. The precise form of this map depends, of course, on the
specific model and its parameters (such as potential strength, etc.).
As we shall see, however, this precise form of $T_E$ is irrelevant for
many qualitative properties of the model such as the spectral type of
$H$. Let us also recall that the {\it Lyapunov exponent} is
defined as
\begin{equation}
\gamma(E)=\lim_{|n|\rightarrow \infty} \frac 1{|n|} \ln \|P_n(E)\|
\end{equation}
The self-similarity properties of substitution sequences can be used to derive
a very efficient method for obtaining crucial information on the asymptotic
properties of $P_n(E)$. Let for any word $\omega\in\cal A^*$
\begin{equation}
T_E(\omega)\equiv \prod_{\alpha\in\omega}T_E(\alpha)
\end{equation}
Set moreover, for $k\in \bf N$,
\begin{equation}
T^{(k)}_E(\omega)\equiv T_E^{(k-1)}(\xi(\omega))
\end{equation}
with $T_E^{(0)}\equiv T_E$. Obviously, the quantities
$T^{(k)}_E(\alpha)$ can then be computed recursively. It has turned
out even more useful to derive from these recursion a system of
recursive equations for the traces of these transfer matrices, called
the {\it trace map}. The existence of a trace map has been established
first in the case of the Fibonacci sequence by Kohmoto et al.
\cite{[8]}, for general substitutions on two-letter alphabets by Allouche
and Peyri\`ere \cite{[43]} and in all generality by Kol\'ar and Nori
\cite{[44]}. More recently, there has been considerable effort by several
groups to find the simplest form of the trace map  \cite{[45],[46],[47]}, but
this is not really relevant for the question we are concerned with. We will
not enter into the derivation of the trace map (for an exposition see e.g.
\cite{[22]}) but only state that in general there exists a finite subset
$\cal B\subset\cal A^*$ such that if we set $x_E^{(k)}(\omega)\equiv
{\hbox {tr}\,} T^{(k)}_E(\omega)$, then there exist for all $\beta\in \cal
B$ polynomial maps $F_\beta: \bf C^{|\cal B|} \rightarrow \bf C$, such that
\begin{equation}
x_E^{(k+1)}(\beta)=F_\beta\left(x_E^{(k)}(\beta_1),\dots,x_E^{(k)}
(\beta_{|\cal B|}) \right)
\end{equation}
In the specific examples that we consider mostly here, the trace map takes the
following form:

\begin{enumerate}
\item (Period doubling sequence) $\cal B=\cal A$, and
\begin{equation}
\begin{array}{ll}
x_E^{(k+1)}(a)&=x_E^{(k)}(a)x_E^{(k)}(b)-2\\
x_E^{(k+1)}(b)&=x_E^{(k)}(a)x_E^{(k)}(a)-2\\
\end{array}
\end{equation}
\item (Thue-Morse sequence) Here it is useful to use slightly more
complicated variables. We set $x^{(k)}_E={\hbox {tr}\,} T^{(k)}_E(a)$ and
$v^{(k)}_E=\frac 12 {\hbox {tr}\,} \left(T^{(k)}_E(a)-T^{(k)}_E(b)\right)^2$
and finally $w^{(k)}_E =v^{(k)}_E+4-\left(x^{(k)}_E\right)^2$. Then \cite{[16]}
\begin{equation}
x^{(k+1)}_E=2-w^{(k)}_E,\quad w^{(k+1)}_E=\left(x^{(k)}_E\right)^2w^{(k)}_E
\end{equation}
\end{enumerate}

These trace maps are universally used tools to investigate these models. They
do depend only on the substitution sequence and {\it not} on the specific
choice of the model or its parameters which enter only through the initial
conditions in a straightforward manner.

To use the trace map in the spectral analysis we need to relate the spectrum
to quantities related to the dynamical system given by the trace map. This
quantity will be, naturally, the {\it stable set } of the dynamical system: we
should expect that a particular value of the energy, $E$, is in the spectrum
of the Hamiltonian, if and only if the corresponding vector of initial traces
$x_E^{(0)}$ remains bounded under subsequent applications of the trace map.

However, this point turns out to be quite subtle and represents, in fact, the
main difficult part in the determination also of the spectral type. First of
all, it is quite easy to give a precise definition of a stable set which
guarantees that it will {\it contain} the spectrum: we say that its
complement, called the unstable set, is the set of initial traces $x_E^{(0)}$
with the property that there exists an $n_0$ large enough such that for all
$n>n_0$, the images under $n$ applications of the trace map have  first
component, $x_E^{(n)}(\alpha_0)$ whose absolute value is larger than two (for a
precise formal definition see e.g. \cite{[22]}). The point is that under this
condition, we can construct a sequence of periodic approximants of $H$,
converging strongly to $H$, for which we know by Floquet theory that $E$ is in
the interior of a spectral gap, and this implies that the same holds for the
limiting operator. Note that this definition does not imply that in the stable
set the traces remain bounded; it suffices that there exists an infinite
number of values $n$ for which the first trace gets smaller than two! The
period doubling sequence furnishes an example in which such behaviour actually
occurs, i.e. for some certain values of the energy the traces undergo
fluctuations with ever growing amplitudes, although, and this is a general
characteristic feature, the growth is slower than exponential (see
\cite{[22]}). This can easily be deduced from the results presented in
\cite{[18]}.

So what about the converse statement? Intuitively, one is tempted to believe
that if the trace of the transfer matrix has modulus less than two, this
should imply that all solutions of both the homogeneous Schr\"odinger equation
(\ref{eq1}) {\it and} the inhomogeneous equation for the Green' s function
\begin{equation}
(H-E)G=\delta_0
\end{equation}
where $\delta_0$ denotes the delta-function concentrated at zero,
cannot tend to zero at infinity and in particular are not
square-summable (recall that the transfer matrices have determinant
one; thus if their trace is less than or equal to two, both their
eigenvalues have modulus one!) which implies first that $E$ is not an
eigenvalue and second that $E$ is in the spectrum of $H$. However, a
moment's reflection will show that this argument is premature and more
work is needed to justify it (the point being that no information is
given on the eigenvectors of the transfer matrices, the angle between which
could tend to zero in which case a decaying solution cannot be excluded).
Still, this argument has been made rigorous in some examples, namely the
Fibonacci sequences, the Thue-Morse sequence and the period-doubling sequence.
However, already in the period-doubling case, this required a fairly
cumbersome analysis that appeared to be impossible to carry through in more
complicated situations.

A less direct, but technically more feasible approach is based on the
following observation: from quite general argument, it is known that the
spectrum of a Schr\"odinger operator must always contain the set of energies
for which the Lyapunov exponent is zero (the converse being false in general;
for random sequences, the Lyapunov is strictly positive even in the spectrum).
Thus, {\it if} one can show that the Lyapunov exponent is zero for all
energies in the stable set, one can close the circle of inclusions and shows
that all three sets are identical! This idea was shown to work in  \cite{[22]}
for a very large class of substitution sequences, namely those giving rise to
so-called ``semi-primitive'' trace maps (for a definition see  \cite{[22]}).

Apart from its generality, an advantage of this method is that it yields as a
by-product immediately the singular nature of the spectrum. By a result of
Kotani \cite{[48]} (augmented by some simple soft analysis that can be found
in  \cite{[18]}) it is known that, for any substitution sequence (excepting
the trivial case of periodic sequences), the set of energies on which the
Lyapunov exponent vanishes must be of zero Lebesgue measure. Thus, the
presence of {\it absolutely continuous spectrum} is immediately excluded in
all these cases! We would like to stress that all these argument apply in the
Kronig-Penney models just as in the tight binding models, even though the
theorems in the original litterature were stated only for the latter case. The
problem left open by this method is that of the possible existence of
eigenvalues. They are excluded under assumptions which allow to show that the
first method works but are certainly too restrictive. It is quite likely that
in fact the spectrum will be singular continuous (i.e. of measure zero and no
eigenvalues) whenever it coincides with the stable set of the trace map, but
some new idea is needed to prove this. The other main open problem is of
course to know whether the assumption of a semi-primitive trace map is really
necessary for the absence of absolutely continuous spectrum. There is one
example, the Rudin-Shapiro sequence, for which the trace map is not
semi-primitive, but unfortunately, there are also no very definite results on
the nature of the spectrum in this case. Dulea et al. \cite{[49]} have
presented some evidence based on a scaling analysis and numerical data that
for strong potentials, the spectrum should be pure point. This would be very
interesting, if it can be confirmed. But of course, all such results should be
regarded with great caution.

\vskip1cm
\section{Extended states and spectra}

We have seen in the last section that the question of the nature of the
spectra in our models is settled rigorously for a large class of substitution
sequences and in particular, with great detail, in the case of the Thue-Morse
sequence. It is still interesting to analyze why and how the authors of
\cite{[30]} arrived at the false conclusion that in this example there exists
an absolutely continuous component in the spectrum. The main basis of their
analysis resides in the distinction of (generalized) eigenstates as `extended'
or `critical'. Now, in the Thue-Morse case (and in fact in all examples where
singular continuity of the spectrum was proven) it turns out that for all
values of the energy in the spectrum, the solutions of the initial value
problem $(H-E)\psi=0$ do not tend to zero at infinity. Thus, reasonably, all
states might be called `extended'.  Still, there are quite different ways a
function that does not tend to zero may behave. A particularly nice one would
be to be periodic, or, at least, to be a possibly aperiodic repetition of
several patterns of constant length. This latter case occurs quite typically
for substitution potentials and it is apparently this behaviour that is
referred to as `extended states'. While this has nothing whatsoever to do with
the presence or absence of absolutely continuous spectrum, this is an
interesting phenomenon and deserves some comments.

Let us for simplicity consider the case of a two-letter alphabet, say
${\cal A}=\{a,b\}$. Now it may happen that for given integer $k$ and for
special
values of the energy $E$ the matrices $T_E^{(k)}(a)$ and $T_E^{(k)}(b)$
commute. Then the two matrices will posses at least one common eigenvector,
and this will give rise to a solution of the Schr\"odinger equation whose
behaviour over steps of length $|\xi^k(a)|$, resp. $|\xi^k(b)|$, we can easily
trace. If one of the traces of the two matrices is larger than two, there are
in fact two such solutions which will grow either to the left or to the right
exponentially fast, so we are out of the spectrum. The interesting case arises
when the two traces have modulus less than or equal to two: in this case they
just multiply the eigenvector by a phase and we obtain an extended solution
whose behaviour is particularly simple. Up to the phase factor, it consists of
two patterns of length $|\xi^k (a)|$ and $|\xi^k (b)|$ which alternate
according to the way the letters $a$ and $b$ are arranged in the substitution
sequence itself.

In general it may be expected that the number of values $E$ for which this
phenomenon occurs should increase with $k$: the entries of the corresponding
matrices are polynomials of ever higher degree in $E$. Of course, the details
here depend on the particular model via the dependence of the basic transfer
matrices on the energy. In the Thue-Morse case and in the period-doubling
case, it has previously been noticed that for each $k$ there exist $2^{k-1}$
values of $E$ for which such solutions exist (see  \cite{[18],[17]} and
\cite{[18]} resp.). Their occurrence in both sequences shows in particular
that they are not tied to the question of quasiperiodicity of the sequence, as
appears to be believed in \cite{[30]}. It should be noted that the case of the
Fibonacci sequence is quite peculiar, as there the matrices $T_E^{(k)}(a)$ and
$T_E^{(k)}(b)$ can only commute if $T_E^{(0)}(a)$ and $T_E^{(0)}(b)$ commute
(this is readily proven by induction; the crucial point is that $A$ commutes
with $AB$ if and only if $A$ commutes with $B$) which, at least in the tight
binding model, is only true in the trivial case where $v\equiv 0$.

In any case, it should be kept in mind that the energies at which such
solutions exist always form at most a countable set. The continuous spectrum,
however, is by its definition uncountable. Therefore, they can never be
significant for absolutely continuous spectrum, and statements like `one half
of the states are extended' are quite meaningless.

We should stress, however, that the properties of generalized eigenfunctions
will be quite important for other physical, in particular transport properties
of these systems (see e.g. \cite{[50],[51]}) even though this entire field is
not yet sufficiently investigated. We would like to mention in this context
that the types of extended states we have discussed above occurs also in
certain cases of random potentials with constraints. An example that was
discussed extensively in recent years is the `random dimer model' (see
\cite{[52],[53]}) introduced to describe certain anomalous conductivity
properties of some polymers.

\vskip1cm
\section{Conclusions}

We have explained in the preceeding sections that for a large class of models
based on substitution sequences the spectrum has no absolutely continuous
component and moreover is concentrated on a set of zero Lebesgue measure.
Moreover, the spectrum coincides in all these cases with the set of energies
for which the Lyapunov exponent is equal to zero (contrary to what is stated
for instance in \cite{[39]}). For some of the most studied examples, namely
the Fibonacci sequences, the Thue-Morse sequence and the period-doubling
sequence, it has moreover been shown that no eigenvalues can exist and
therefore the spectrum is purely singular continuous in these cases. There
appears to be no connection between this fact and the nature of the Fourier
spectrum of the sequences.

There are a number of important problems that remain, however, open. First, we
would like to see whether the absence of absolutely continuous spectrum can be
proven for {\it all} aperiodic substitution sequences. We expect this to be
true. Second, it would be interesting to see whether there are examples of
substitution sequences which give rise to point spectrum. A candidate is here
the Rudin-Shapiro sequence and it is well worth-while to study it more
thoroughly.

Further questions regard the more detailed structure of the spectra, beyond
just the spectral type. Only in the Thue-Morse and the period-doubling cases
is a very detailed description of the spectrum available (e.g. precise
asymptotics for the behaviour of all spectral gaps). It is a quite bothering
feature that even in the simplest case of all, the golden Fibonacci sequence,
the opening of the gaps at small potentials is not known! In this context it
would be very interesting to give a more rigorous foundation to the
perturbation theoretic arguments of Luck \cite{[19]}. Also one would like to
see more precise relations between characteristics of the spectra (e.g.
fractal and correlation dimensions of the spectral measure) and actual
transport properties. At least from a mathematical point of view, there is
little more than some vague ideas that exist so far, and we do not want to
discuss this point here.

Finally, all results so far are obtained for models that are associated to
products of two-by-two matrices (excepting the gap-labelling theorems
\cite{[20],[21]} which are valid for much larger classes of systems). Thus
models with long-range hopping models on a strip are not covered by existing
theorems. The reason for this is basically that there is no analogue of the
trace map known in these cases. This leaves a lot of room for further
investigations.

\parskip1pt plus 3pt


\end{document}